\documentclass{PoS}

\title{Hard fast X-ray transients as possible counterparts of unidentified MeV/TeV sources}

\ShortTitle{Fast X-ray transients as counterparts of unidentified MeV/TeV sources}

\author{\speaker{V. Sguera}\thanks{I am very grateful to L. Bassani, A. Bazzano, A.J. Bird and G.E. Romero for their collaboration.
This research made use of data obtained from the HEASARC database.}\\
        INAF/IASF Bologna, via Piero Gobetti 101, 40129 Bologna, Italy\\
        E-mail: \email{sguera@iasfbo.inaf.it}\\

}

\abstract{The two still unidentified MeV sources EGR J1122$-$5946 and AGL J2022+3622 are here tentatively associated 
with soft gamma-ray candidate counterparts detected through INTEGRAL/IBIS observations. 
On the basis of spatial proximity and/or similar transient behaviour, 
we propose the supergiant fast X-ray transient (SFXT) IGR J11215$-$5952 and the 
candidate SFXT IGR J20188+3647 as possible candidate counterparts of EGR J1122$-$5946 and AGL J2022+3622, respectively.
Our findings possibly suggest that hard fast X-ray transients could 
represent a new class of galactic transient MeV/TeV emitters. Additional evidence 
for the existence of such new class is also provided by very recent AGILE and GLAST discoveries on 
the galactic plane of several unidentified transient MeV sources lasting only a few days}

\FullConference{7th INTEGRAL Workshop\\
		 September 8-11 2008\\
		 Copenhagen, Denmark}

\begin{document}

\section{Introduction}
The field of high energy astronomy is relatively new and young. Breakthrough results have been obtained 
in the last twenty years thanks to gamma-ray satellites 
carrying instruments such as CGRO/EGRET, INTEGRAL/IBIS and Swift/BAT whose survey capabilities 
unveiled the extreme richness of objects in the gamma-ray sky (Hartman et al. 1999, Bird et al. 2007, Casandjian \& Grenier 2008).
Recently, ground-based TeV astronomy  has also shown rapid progress 
with important results reported by the  third generation of imaging atmospheric Cherenkov telescopes 
such as HESS, MAGIC, VERITAS and CANGAROO. A rapidly growing list of $\sim$ 75 TeV sources have been 
detected, $\sim$35 \% of which are still unidentified (Ribo 2008) 

Among the different types of sources shining in the high energy sky, 
High Mass X-ray Binaries (HMXBs) rapidly became the subject of topical and major 
interest in high energy astronomy.  During the last few years, four HMXBs have been detected
at TeV energies, namely Cygnus X-1 (Albert et al. 2007),  LS 5039 (Aharonian et al. 2005a), LS I+61 303 (Albert et al. 2006)
and PSR B1259$-$63 (Aharonian et al. 2005b); their TeV emission provides evidence
that particles can be efficiently accelerated to very high energies.
Furthermore, in the last few years the IBIS instrument (Ubertini et al. 2003) 
on board the INTEGRAL satellite (Winkler et al. 2003) has revolutionised our 
classical view of HMXBs previously built over $\sim$ 40 years of X-ray observations. 
Among the others, IBIS has discovered  many supergiant HMXBs (SGXBs) with X-ray characteristics never seen before; 
they represent a new class of X-ray binaries which have been named as
Supergiant Fast X-ray Transients, SFXTs (Sguera et al. 2005,2006, Negueruela et al. 2006). 
SFXTs spend most of the time in a low level of X-ray activity characterized by luminosities in the range 
10$^{32}$--10$^{34}$ erg s$^{-1}$, well below the persistent state of classical SGXBs ($\sim$10$^{36}$ erg s$^{-1}$).
Only occasionally they do display fast X-ray flares, lasting typically a few hours, with a dinamical range of 
$\sim$10$^{3}$--10$^{4}$. Two main models have been proposed to explain 
their peculiar fast X-ray flaring behaviour, they invoke spherically simmetric clumpy wind
from the supergiant companion (Negueruela et al. 2008, Walter \& Zurita 2007) and anisotropic wind (Sidoli et al. 2007), respectively.

A common characteristic of the IBIS and EGRET catalogs (Hartmann et al. 1997, Bird et al. 2007, Casandjian \& Grenier 2008)
is that a considerable fraction of their high energy sources is still unidentified, with no counterpart at other wavelengths.
This indicates the need for continuing their studies with current missions such as  INTEGRAL, GLAST, AGILE, in order
to find counterparts capable to produce this high energy radiation. To this aim, very recently Sguera et al. (2008 submitted)
proposed the SFXT AX J1841.0$-$0536 as a fast transient MeV/TeV emitter, based on spatial proximity and similar X-ray/soft $\gamma$-ray behavior
with the unidentified sources 3EG J1837$-$0423 and HESS J1841$-$055. Such association is also supported 
from an energetic standpoint by a theoretical scenario where AX J1841.0$-$0536 is a low magnetized 
pulsar which, due to accretion of a massive clump from its supergiant companion donor, undergoes sporadic changes to a transient 
state where a magnetic tower can produce transient jets and as a consequence  high energy emission (Sguera et al. 2008 submitted).
Based on their findings, the authors proposed the SFXT AX J1841.0$-$0536 as the prototype of a new class of
galactic fast transient MeV/TeV emitters. Additional evidences  for the existence of such new class is also provided by very 
recent AGILE and GLAST discoveries on  the galactic plane of several unidentified transient 
MeV sources lasting only a few days (Chen et al. 2007, Pittori et al. 2008, Longo et al. 2008, Cheung et al. 2008).

Here we propose two other fast X-ray transients, IGR J11215$-$5952 and IGR J20188+3647,
as best candidate counterparts of two unidentified MeV sources,
EGR J1122$-$5946 and AGL J2022+3622 respectively. 
Our proposed associations are mainly based on spatial correlation  and/or similar temporal behaviour.

\section{AGL J2022+3622}
The transient MeV source AGL J2022+3622 (E$>$100 MeV) was discovered in the Cygnus region by
AGILE on 23 November 2007 (Chen et al. 2007). Its positional error circle  is centered at galactic coordinates l=75$^\circ$.0,
b=--0$^\circ$.4 with an error radius of $\sim$ 1$^\circ$. AGL J2022+3622 was characterized by a 
strongly variable gamma-ray emission lasting  only one day. 

\begin{figure}[t!]
\includegraphics[width=16cm,height=9cm]{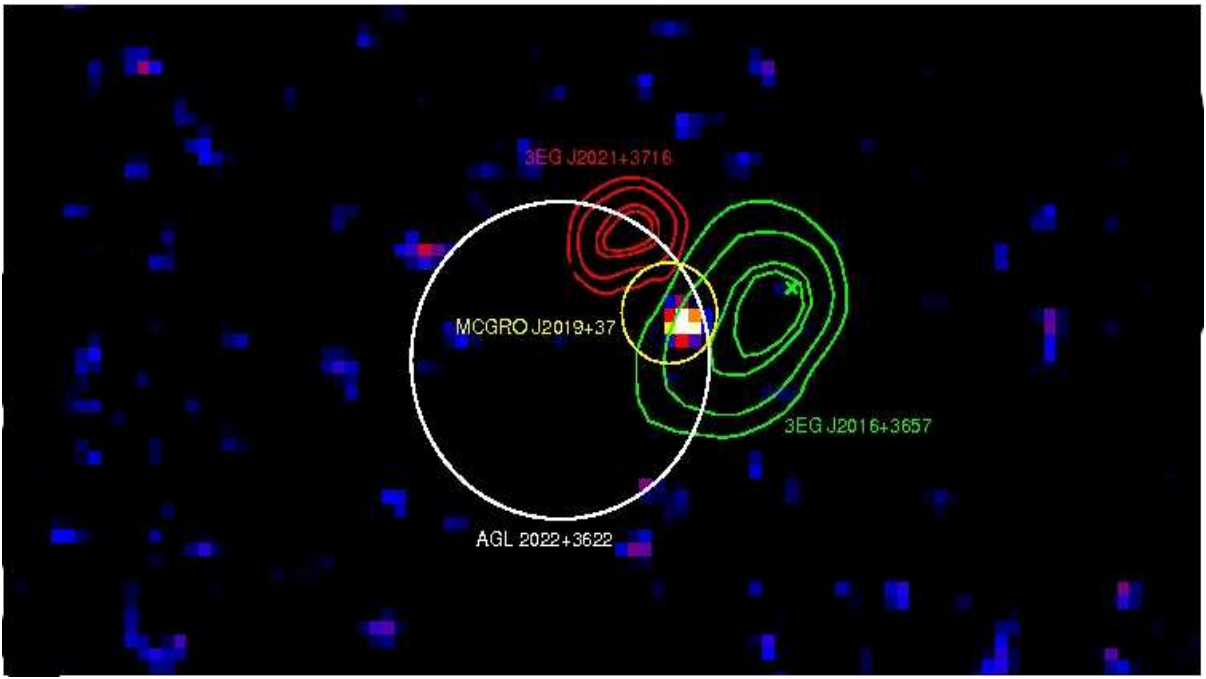}
\caption{IBIS 17--30 keV significance image ($\sim$ 1 hour exposure) of the fast hard X-ray transient 
IGR J20188+3647 ($\sim$ 7$\sigma$ detection) with superimposed: AGL J2022+3622 error circle in white, 
MGRO J2019+37 error circle in yellow, 3EG J2016+3657 probability contours in green (from 50\% to 99\%) and 3EG J2021+3716 
probability contours in red (from 50\% to 99\%).
}
\end{figure}
\begin{figure}[t!]
\includegraphics[width=15.5cm,height=8.5cm]{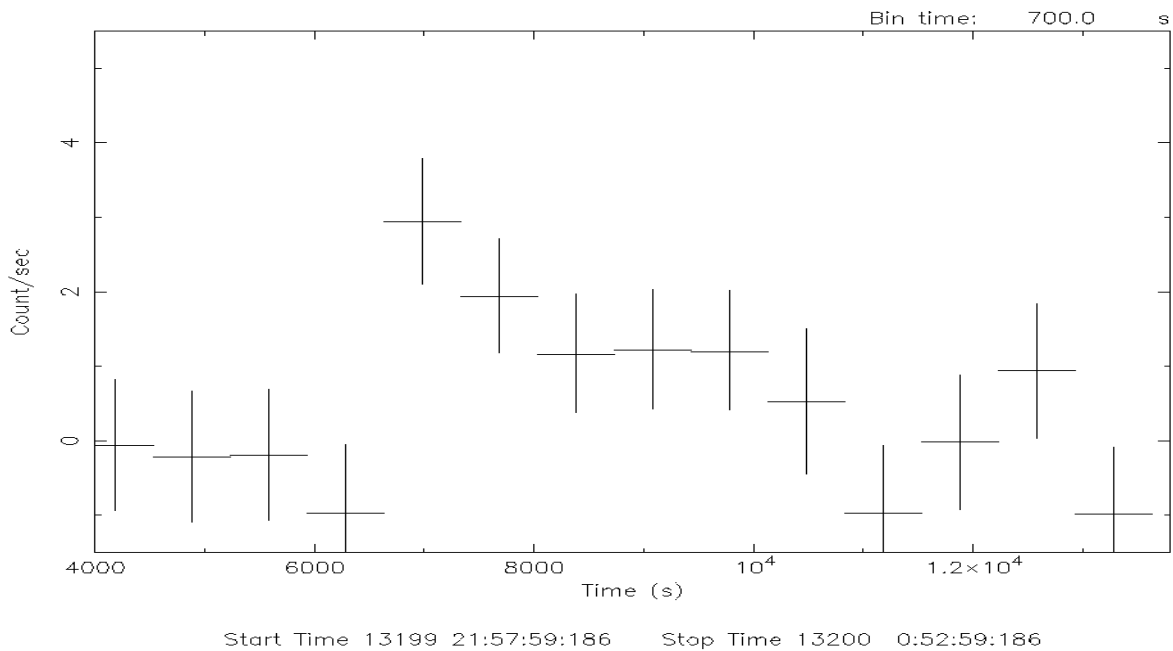}
\caption{ISGRI 17--30 keV light curve of the fast hard X-ray transient IGR J20188+3647 pertaining to its discovery by IBIS.}
\end{figure}

\subsection{IBIS observations of AGL 2022+3622}

Unfortunately, no simultaneous IBIS observations of the AGILE source's discovery are available. 
The region of the sky including AGL J2022+3622 was observed by IBIS $\sim$ 9 days after the AGILE discovery for 1 $\sim$ Ms 
but no hard X-ray source was detected inside the AGILE error circle, 
providing  a  2$\sigma$ upper limit of $\sim$ 0.4 mCrab (20--40 keV).
To date, the region of the sky including AGL J2022+3622 has been observed  by IBIS for a total of  $\sim$ 2 Ms.
Only in one occasion, on  13 July 2004, IBIS detected a newly discovered fast hard X-ray transient source
that is located inside the AGILE error circle: IGR J20188+3647 (Sguera et al. 2006b). 
This source is clearly detected at $\sim$ 7$\sigma$ level (17--30 keV) in the IBIS significance image 
shown in  fig. 1 ($\sim$ 1 hour exposure time) at coordinates RA=20 18 48, Dec=36 47 31.2,
error radius= 3.4 arcminutes (90\% confidence). The average flux is equal to $\sim$ 33 mCrab or 
2.1$\times$10$^{-10}$ erg cm$^{-2}$ s$^{-1}$ (17--30 keV). We note that the poorly known and unidentified EXOSAT source 
EXMS B2016+366 (Reynolds et al. 1999)  is located inside the error circle of IGR J20188+3647. Furthermore, IGR J20188+3647
is located inside the error box of the poorly known and unidentified HEAO A-1 source 1H2018+366 (Wood K.S. et al. 1984). 
It could be that the three hard X-ray sources are the same object. Fig. 1 clearly shows that the Agile source 
AGL J2022+3622  and IGR J20188+3647 are spatially correlated.  IGR J20188+3647 is the only hard X-ray source (E$>$10 keV)
located inside the AGILE  positional uncertainty.

Although the statistics are not good enough to perform a detailed spectral analysis, we were able to 
extract an ISGRI spectrum of  IGR J20188+3647 (17--30 keV) from the unique Science Window (ScW)
during which it was detected ($\sim$ 1 hour exposure time). A power law ($\chi^{2}_{\nu}$=1.6, d.o.f 3), 
or alternatively a thermal bremsstrahlung ($\chi^{2}_{\nu}$=1.7, d.o.f 3), 
gave reasonable fits although it was not possible to fully constrain the spectral parameters
($\Gamma$ $\sim$ 2.4 and kT $\sim$ 14 keV, respectively).
Fig. 2 shows the ISGRI light curve  (17--30 keV) of IGR J20188+3647, it is evident the transient behaviour
with a fast rise ($\sim$ 10 minutes) followed by a slower decay ($\sim$ 50 minutes). 
This kind of fast X-ray flaring activity  and spectral parameters strongly resemble those of SFXTs and 
we can consider it a candidate SFXT.

It is worth noting that IGR J20188+3647 and AGL J2022+3622 are located in a region of the sky rich 
of other high energy sources, as can be clearly seen in fig.1. Specifically, we highlight the presence of the 
following gamma-ray sources: 
\begin{itemize}
  \item  MGRO J2019+37 (smaller and yellow circle in fig. 1). It is an unidentified and highly extended TeV source discovered by 
the water Cherenkov detector Milagro (Abdo et al. 2007). Although fig.1 clearly shows that MGRO J2019+37 is spatially correlated 
  with IGR J20188+3647, the extended and diffuse nature of the TeV source seems to exclude a physical association. 
  \item 3EG J2016+3657 (green probability contours at 50\%, 68\%, 95\%  and 99\%). 
  It has been optically identified by Halpern et al. (2001)  with the blazar B2013+370 indicated by the green x point  in fig. 1, 
  so we can rule out a physical association between 3EG J2016+3657 and IGR J20188+3647.  
  \item  3EG J2021+3716 (red probability contours at 50\%, 68\%, 95\%  and 99\%). Very recently, AGILE observations (E$>$100 MeV)
  identified its counterpart with the pulsar PSR J2021+3651 (Halpern et al. 2008)  that does not show variability at gamma-rays.
\end{itemize}
It is worth pointing out that the above EGRET sources are from the third EGRET catalog (Hartman et al. 1997). Comparing 
with the very recent and revised EGRET catalog by Casandjian \& Grenier (2008), we note that 
3EG J2016+3657  is not listed anymore while  3EG J2021+3716 is still listed with coordinates
similar to those from the third EGRET catalog.

In the light of all the findings reported above, we propose the candidate SFXT IGR J20188+3647 
as one possible candidate counterpart of the fast transient gamma-ray source AGL J2022+3622, to date. 
Further and deep studies of the entire region in X-rays (i.e. XMM, Chandra and Swift/XRT), soft gamma-rays (i.e. INTEGRAL), MeV and GeV 
(i.e. AGILE and GLAST) are strongly needed in order to confirm or reject our proposed identification.

\section{EGR J1122$-$5946}

Recently, the whole EGRET dataset (E$>$100 MeV) has been reanalyzed by using a new and much improved galactic interstellar 
emission model based on very recent dark gas, CO, HI, and interstellar radiation field data. 
The results have been reported by Casandjian \& Grenier (2008) in a revised EGRET catalog which lists 188 
sources compared to the 271 entries of the previous third EGRET catalog.
About 107 former EGRET sources have not been confirmed  because of structures in the interstellar background, the majority of them 
were unidentified sources associated with the local clouds of the Gould belt.
It is worth pointing out that the revised catalog by Casandjian \& Grenier (2008) also lists 30 new gamma-ray sources not previously 
reported in the third EGRET catalog. Among them, we focussed our attention on the newly discovered and
unidentified gamma-ray source EGR J1122$-$5946. It is located at RA=170$^\circ$.55 and Dec=-59$^\circ$.77 with an
error circle radius of 0$^\circ$.31. Unfortunately, it is still unknown if EGR J1122$-$5946 is a persistent or a transient 
gamma-ray object since no such information is provided in the revised catalog. 
However from the reported fluxes (E$>$100 MeV) in three different EGRET viewing period lasting $\sim$ one year each,
the source seems to be slightly variable  (F$_{P1}$(Apr1991-Nov1992)=27$\pm$7$\times$10$^{-8}$  photon cm$^{-2}$ s$^{-1}$, F$_{P3}$(Aug1993-Oct1994)=9$\pm$8$\times$10$^{-8}$ 
photon cm$^{-2}$ s$^{-1}$,  F$_{P4}$(Oct1994-Oct1995)=53$\pm$15$\times$10$^{-8}$ 
photon cm$^{-2}$ s$^{-1}$). 

\subsection{IBIS observations of EGR J1122$-$5946}

The region of the sky including EGR J1122$-$5946 has been extensively covered by IBIS
observations and we exploited them with the aim of finding its best candidate counterpart.
Fig. 3 shows the 20--100 keV IBIS significance mosaic ($\sim$ 2.5 Ms exposure) superimposed on 95\% EGRET error circle.
We note that no persistent soft gamma-ray source has been detected inside of it despite the deep IBIS observation. 
As next step, we took into account the possibility that a fast X-ray transient source, active on short timescale (i.e. few hours),
could be located inside the error circle of
EGR J1122$-$5946 and it  was not detected in the deep IBIS mosaic because 
the integration on long exposure time  degrades the signal to noise to very low values.
Bearing such possibility in mind, we performed an analysis at ScW level ($\sim$ 2 ks) of the entire IBIS 
ScW dataset ($\sim$ 2.5 Ms). By doing so  we unveiled the presence of 
the SFXT IGR J11215$-$5952 inside the EGR J1122$-$5946 error circle. This can be clearly seen in fig. 4 
which shows the 20--100 keV IBIS significance mosaic taking into account only the ScWs during which 
IGR J11215$-$5952 was active ($\sim$ 45 ks exposure time). No other catalogued hard X-ray sources (E$>$10 keV) are located
inside the  EGR J1122$-$5946 error circle, according with all the available catalogs from the HEASARC database.

IGR J11215$-$5952 is a HMXB binary system hosting a $\sim$ 187 seconds neutron star as compact object 
and a massive supergiant star as companion donor. It is characterized by a fast X-ray transient 
behaviour (Sidoli et al. 2006,2007, Romano et al. 2007) and classified as SFXT.
The orbit of IGR J11215$-$5952 is highly eccentric and the neutron star 
spends most of the time far away from the companion donor; the typical quiescent luminosity 
of the system is L$_x$ $\sim$ 10$^{33}$ erg s$^{-1}$. Accretion takes place only at periastron passage (every $\sim$ 165 days), 
when the compact object approaches the supergiant star; the brightest part of the outburst activity 
consists of fast X-ray flares ($\sim$ hours) with a typical X-ray luminosity of 10$^{36}$ erg s$^{-1}$.

As we can clearly see in fig. 4, IGR J11215$-$5952 is spatially correlated with the unidentified EGRET source EGR J1122$-$5946
and this makes tempting to postulate a physical correlation between the two objects. Unfortunately,  
informations on the temporal behaviour of EGR J1122$-$5946 (i.e. persistent or transient nature), which could shed more light 
on its association with the SFXT, are still unknown and a further study from this point of view is strongly needed.
Finally, we took into account the possibility that the spatial association could be simply a chance coincidence.
To this aim,  we calculated the probability of finding a supergiant HMXB, such as IGR J11215$-$5952, 
inside the  EGR J1122$-$5946 error circle ($\sim$ 0$^\circ$.3) by chance. 
Given the number of  supergiant HMXBs detected by IBIS within the galactic plane (Bird et al. 2007), defined here as 
a latitude range $\pm$3$^\circ$, we estimated a probability of $\sim$0.6\%, i.e. low enough to claim a physical
association between EGR J1122$-$5946 and IGR J11215$-$5952.

\begin{figure}[h!]
\includegraphics[width=16cm,height=8.5cm]{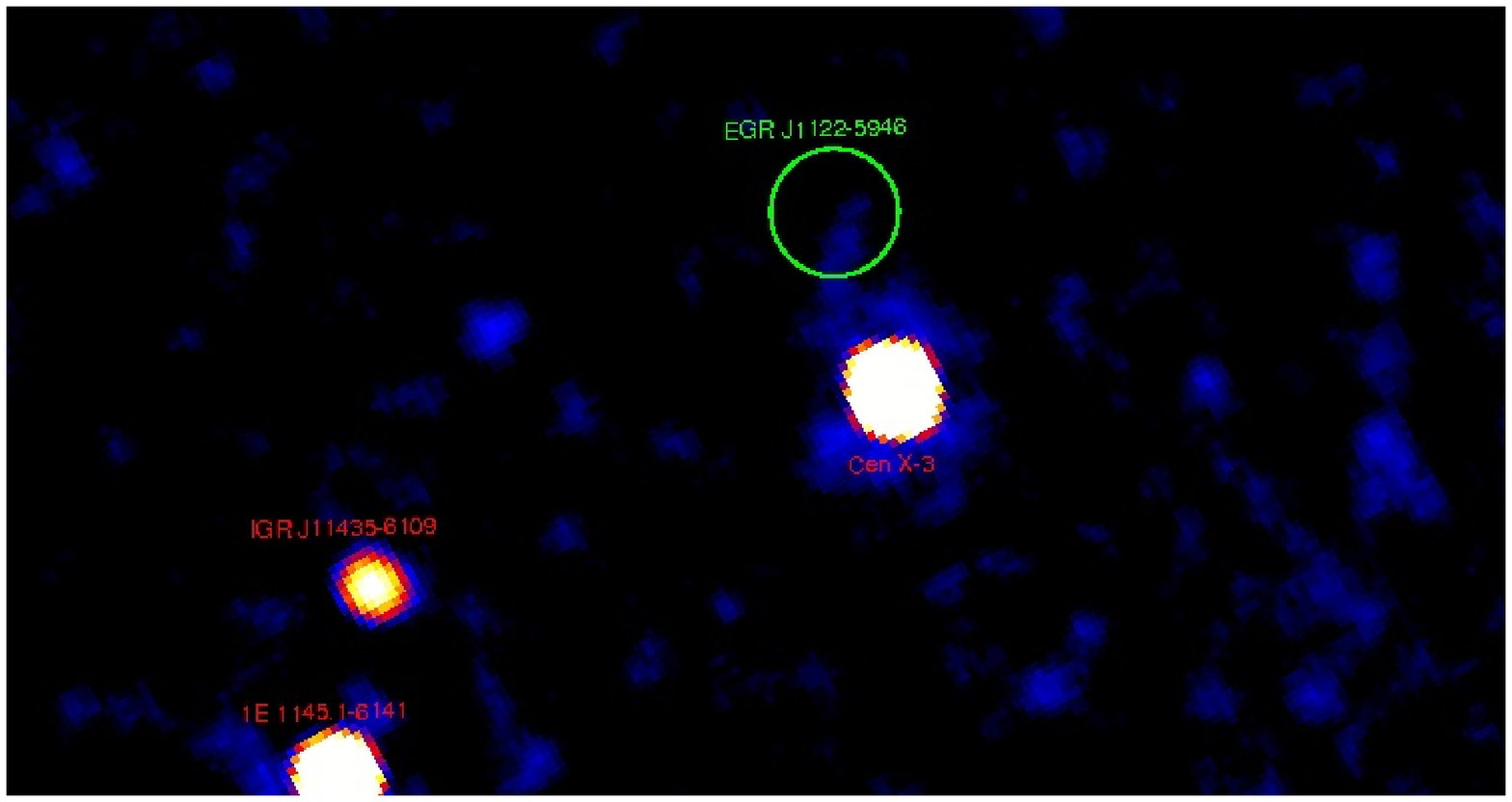}
\caption{IBIS 20--100 keV significance deep mosaic image ($\sim$ 2.5 Ms) superimposed on the EGR J1122$-$5946 error circle.
We note that no persistent soft gamma-ray source has been detected inside the EGRET error circle.}
\includegraphics[width=16cm,height=8.5cm]{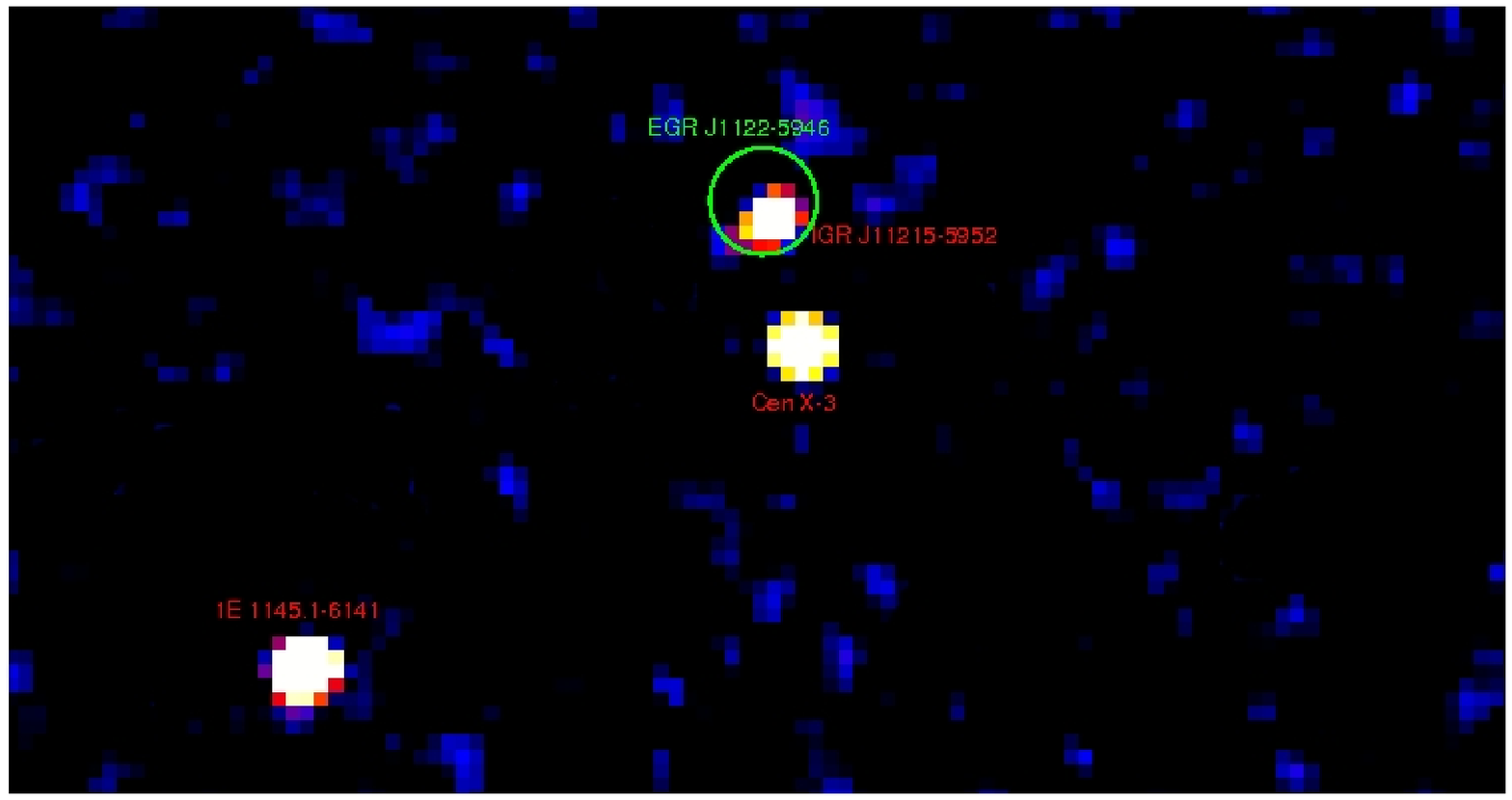}
\caption{IBIS 20--100 keV significance mosaic image ($\sim$ 45 ks) taking into account only the ScWs during which the 
SFXT IGR J11215$-$5952 was active. It is evident the spatial correlation between the SFXT and the unidentified source 
EGR J1122$-$5946 (green error circle). Contrary to Fig. 3, we note that the Be HMXB IGR J11435$-$6109 has not been detected because 
it was not active during such very short exposure time.}
\end{figure}

\section{Discussion and future work}
In the light of the findings previously reported, we propose the SFXT IGR J11215$-$5952 and the candidate SFXT 
IGR J20188+3647 as possible candidate counterparts of the unidentified  MeV sources EGR J1122$-$5946 and AGL J2022+3622, respectively.
It is worth pointing out that the possible fast MeV flares from IGR J11215$-$5952 and IGR J20188+3647
might not be a unique and rare case.  Another SFXT, AX J1841.0$-$0536, has 
also been proposed as a fast transient MeV/TeV emitter based on spatial proximity and similar X-ray/soft $\gamma$-ray behavior
with the unidentified sources 3EG J1837$-$0423 and HESS J1841$-$055 (Sguera et al. 2008 submitted). Additional evidences 
for the existence of such a new class of  galactic fast transient MeV/TeV emitters  is also provided by very 
recent AGILE and GLAST discoveries on  the galactic plane of several unidentified transient 
sources lasting only a few days (Chen et al. 2007, Pittori et al. 2008, Longo et al. 2008,Cheung et al. 2008).
We stress that flaring high energy emission could be common among HMXBs, this is supported by the fact that
recently the supergiant HMXB Cygnus X-1 has been detected at TeV energies during a 
short flare lasting $\sim$ 80 minutes (Albert et al. 2007). Interestingly, such TeV flare was simultaneous to a 
20--40 keV hard X-ray flare detected by IBIS at $\sim$ 1.5 Crab level (Turler et al. 2006). 
Short TeV flares have been also detected in two other HMXBs, LS 5039 (De Naurois 2006, Paredes 2008) and LS I +61 303 (Paredes 2008).

As for the physical mechanism responsible for the fast gamma-ray flares, it is important to point out that 
four HMXBs and no LMXBS have been detected at MeV/TeV energies, to date. This fact points to the 
need of having a bright and massive OB star as the source of seed photons (for the Inverse Compton emission) 
and target nuclei (for hadronic interactions). 
Sguera et al. (2008 submitted) proposed a theoretical mechanism able to explain the flaring MeV emission 
from  the SFXT AX J1841.0$-$0536. In this scenario,  AX J1841.0$-$0536 is a low magnetized pulsar 
which,  due to accretion of a massive clump of material from the supergiant companion, undergoes sporadic changes to a
transient Atoll-state where a magnetic tower  can produce transient jets and as a consequence high energy transient emission. 
After the collision with the massive clump, everything comes back to the normal state.
Moreover, Walter (2007b) suggested that HMXBs accreting dense clumps of material from the stellar wind
could be transient MeV/TeV sources if the column density is large enough
i.e. $\ge$ $\sim$ 10$^{23}$ cm$^{-2}$. Since such column densities have been observed in specific SFXTs during fast X-ray flares,
the detection of fast MeV/TeV flares on a few hours timescale it is not unexpected.
Specifically, protons trapped in the outer and closed regions of a neutron star
magnetosphere could be accelerated up to a Lorentz factor of $\sim$10$^8$ by multiple scattering of Alfv\'en 
waves in or close to the accretion column and then interact with dense clump of materials at the magnetospheric
radius producing $\gamma$-rays through inelastic $pp$ collisions and the subsequent decays.
The duration and strength of the $\gamma$-rays flares should depend on 
the size of the clumps, structure of the wind and magnetic strength. 

In summary, our results show that hard fast X-ray transients
could represent a new class of galactic MeV/TeV emitters. The high energy flaring emission from the binary system 
is expected only for a very small fraction of time so that fast  MeV/TeV flares are not easy to detect.
Further multiwavelength observations and deeper studies in radio, X-rays (i.e. XMM, Chandra and Swift/XRT), soft gamma-rays (i.e. INTEGRAL), 
MeV and GeV frequencies (i.e. AGILE and GLAST) are strongly needed in order to support or reject our proposed scenario.

\end{document}